\documentstyle[epsfig]{ioplppt}                
\begin{document}
\jl{4}
\title{Test of Hadronic Interaction Models in the Forward Region
with KASCADE Event Rates }

\author{
T~Antoni\dag, W~D~Apel\dag,
F~Badea\ddag, K~Bekk\dag, 
A~Bercuci\ddag, \\
K~Bernl\"ohr\dag\ftnote{6}
{Now at: Humboldt University, 10099 Berlin, Germany},
H~Bl\"umer\dag\S, E~Bollmann\dag, 
H~Bozdog\ddag, I~M~Brancus\ddag, C~B\"uttner\dag, 
A~Chilingarian\#, K~Daumiller\S, P~Doll\dag, 
J~Engler\dag, F~Fe{\ss}ler\dag, H~J~Gils\dag, R~Glasstetter\S, 
R~Haeusler\dag,
A~Haungs\dag, D~Heck\dag, J~R~H\"orandel\S, 
T~Holst\dag, A~Iwan*,
K-H~Kampert\dag\S, J~Kempa*\ftnote{8}{Now at: Warsaw University of
Technology, 09--400 Plock, Poland},
H~O~Klages\dag, 
J~Knapp\S\ftnote{9}{Now at: University of Leeds, Leeds LS2 9JT, U.K.},
G~Maier\dag, H~J~Mathes\dag, 
H~J~Mayer\dag, J~Milke\dag, 
M~M\"uller\dag, 
J~Oehlschl\"ager\dag,  \\
S~S~Ostapchenko\dag\ftnote{4}{On leave of absence from the
 Moscow State University, Moscow, Russia},
M~Petcu\ddag, H~Rebel\dag, 
M~Risse\dag\ftnote{5}{Corresponding author. FAX: +49--7247--82--4047,
                   E--mail: risse@ik1.fzk.de},
M~Roth\dag, G~Schatz\dag, 
H~Schieler\dag, 
J~Scholz\dag,
T~Thouw\dag, H~Ulrich\dag, 
J~Unger\dag,
B~Vulpescu\ddag,
J~H~Weber\S, J~Wentz\dag, 
J~Wochele\dag, 
J~Zabierowski**, S~Zagromski\dag}
\address{\dag\ Institut f\"ur Kernphysik, Forschungszentrum Karlsruhe,
D--76021 Karlsruhe, Germany}
\address{\ddag\  National Institute of Physics and Nuclear Engineering,
P.O. Box Mg--6, RO--7690 Bucharest, Romania}
\address{\S\ Institut f\"ur Experimentelle Kernphysik, Universit\"at
Karlsruhe, D--76021 Karlsruhe, Germany}
\address{\# Cosmic Ray Division, Yerevan Physics Institute, Yerevan
36, Armenia}
\address{* Department of Experimental Physics, University of Lodz, 
PL--90236 Lodz, Poland}
\address{** Soltan Institute for Nuclear Studies, PL--90950 Lodz, Poland}

\begin{abstract}
An analysis of muon and hadron rates observed in the central detector
of the KASCADE experiment has been carried out.
The data are compared to CORSIKA simulations employing the
high-energy hadronic interaction models QGSJET, DPMJET, HDPM, SIBYLL, 
and VENUS.
In addition, first results with the new hadronic interaction model
\mbox{neXus 2} are discussed.
Differences of the
model predictions, both among each other and when confronted with
measurements, are observed. 
The hadron rates mainly depend on the inelastic
cross-section and on the contribution of
diffraction dissociation. The discrepancy between simulations and
measurements at low primary
energies around 5 TeV can be reduced by increasing the non-diffractive
part of the inelastic cross-section of nucleon-air interactions.
Examination of hadron multiplicities points towards harder spectra of
secondary pions and kaons needed in the calculations.
\end{abstract}

{\bf To be published in Journal of Physics G}

\section{Proem}

The observation of extensive air showers (EAS) provides an opportunity
to study the hadronic interaction in an energy range exceeding that
of present artificial accelerators, and moreover in
kinematical ranges essentially unexplored in collider experiments.
The diffractive particle production, which strongly influences the EAS
development and energy flux in forward direction \cite{jones}, has been
experimentally investigated at comparatively low energies
(\mbox{$\sqrt{s} \simeq $ 10 GeV}, for a review see \cite{kaidalov}).
At higher energies, the
UA5 experiment \mbox{($\sqrt{s} = 0.9$ TeV)} could register
only \mbox{30 \%} and the CDF detector
(\mbox{$\sqrt{s}= 1.8$ TeV}) even only about \mbox{5 \%}
of the total energy.
Additionally one should keep in mind that these experiments examined
high-energy proton-antiproton collisions, while nitrogen is the most
abundant target nucleus in EAS.
Hence, hadronic interaction models for the higher energy regime rely on
extrapolations from lower energies, guided by more or less detailed
theoretical (QCD inspired) prescriptions. 

The resulting models
of the hadronic interactions suffer from various uncertainties,
firstly due to necessary approximations done in the model construction,
secondly because of systematic uncertainties and inconsistencies
in the experimental results basing the extrapolations.
For example, nucleon-carbon cross-sections have been measured at energies of
\mbox{200$-$280 GeV} as $\sigma_{inel} =$
\mbox{225$\pm$7 mb \cite{carrol}} and
\mbox{237$\pm$2 mb} \cite{roberts} with uncertainties mainly of
systematical character. At the highest energy of
\mbox{$\sqrt{s} = 1.8$ TeV},
corresponding to $E_{lab} \simeq 1.7$ PeV, results obtained for
the total proton-antiproton cross-section are $\sigma_{tot} =$
\mbox{72.8$\pm$3.1 mb}
(E710 \cite{E710}), \mbox{80.03$\pm$2.24 mb} (CDF \cite{CDF}),
and \mbox{71.71$\pm$2.02 mb} (E811 \cite{E811})
with a probability of the values being consistent with each other
of only \mbox{1.6 \%} \cite{E811}.
These systematic uncertainties of \mbox{5$-$10 \%} are propagated when
constructing hadron-air and nucleus-air cross-sections for EAS
simulations. 

The effects on the EAS predictions are considerable.
As EAS simulations using the Monte Carlo program CORSIKA \cite{heck}
show, for proton primaries of \mbox{$10^{14}$$-$$10^{15}$ eV}
an increase of $\sigma_{inel}$(hadron-air) by \mbox{10 \%}, e.g.,
leads to a reduction of the number 
of high-energy hadrons \mbox{($>$100 GeV)} by up to \mbox{50 \%}
and of the electron
number \mbox{($>$3 MeV)} by \mbox{$\simeq$15 \%}, calculated for
EAS of vertical incidence at sea level.
The depth of shower maximum gets shifted higher in the atmosphere by
about \mbox{15 g/cm$^2$}.
The total muon number \mbox{($>$300 MeV)} is relatively little
affected (\mbox{$\simeq$4 \%}), while the lateral distribution appears
to be considerably flatter with increasing $\sigma_{inel}$.

These model uncertainties also influence astrophysical interpretations
of EAS measurements. A salient
example is the mass composition at the
highest primary energies deduced from the depth of shower maximum. The
composition varies from mixed to pure iron depending on the
model (see, e.g., reference \cite{watson}).

Though studies of high-energy hadronic interactions by EAS 
observations imply a limited knowledge about the nature of the 
interacting primary, the strong dependence of
the hadronic EAS observables on details of the hadronic interaction
models enables a test 
with effective constraints on the model parameters.
The central detector of the KASCADE experiment \cite{klages, engler}
with a large hadron calorimeter and muon detectors provides
unique experimental possibilities for such an endeavour.
A recent analysis of the hadronic structure in EAS
cores at primary energies around the {\it knee} \cite{joerg},
i.e.~in the PeV region,
 exhibited differences
between the models QGSJET \cite{qgsjet}, VENUS \cite{venus},
and \mbox{SIBYLL \cite{sibyll}}. In the
present article we report on tests at lower primary energies
of \mbox{0.5$-$500 TeV}. At these energies, 
\begin{itemize}
 \item the models should be quite reliable due to some overlap with
       accelerator data and
 \item the primary flux and composition are reasonably well known from
 balloon and satellite borne experiments, i.e.~with an uncertainty of
 about \mbox{15 \%} at \mbox{10 TeV
 \cite{watson,wiebel}}.
\end{itemize}

The idea is to measure the rate
of events in the central detector, when a certain number of
trigger counters (mostly induced by muons) have fired and
at least one hadron has been detected in the calorimeter.
These rates prove to be a stringent test quantity. They are
compared with predictions of EAS simulations generated by the
models QGSJET, SIBYLL, VENUS, \mbox{DPMJET} \cite{ranft},
\mbox{HDPM} \cite{heck,capde},
and \mbox{{\sc \normalsize neXus} \cite{drescher}}.

\section{Experimental concept}

\begin{figure}\centering
   \epsfig{file=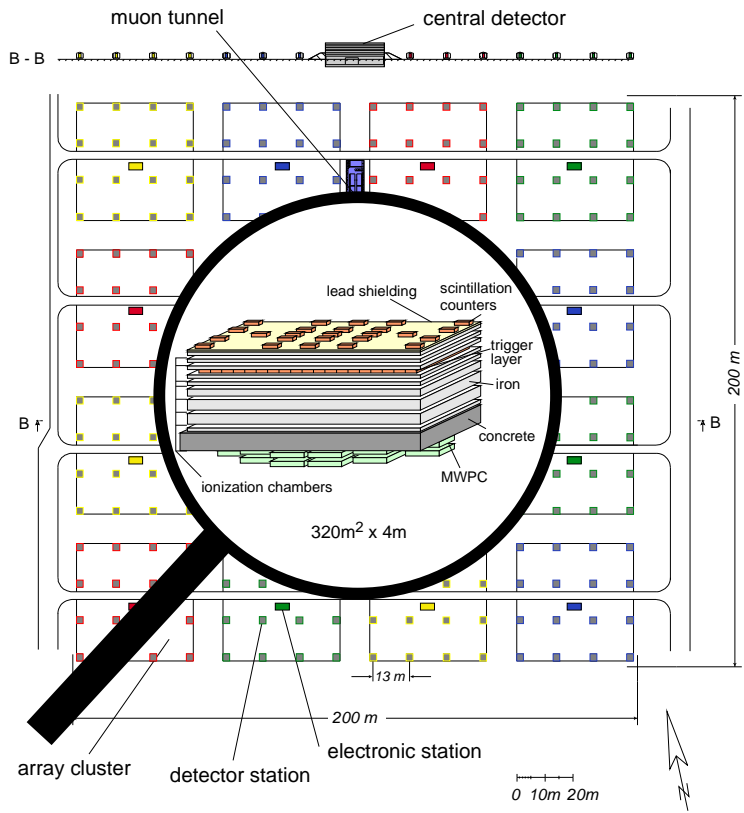,width=0.90\textwidth}
   \caption{Arrangement of the KASCADE installation. In the array
   stations muons (except in the inner four clusters)
   and the electromagnetic component are measured.
   Enlarged in the centre the hadron calorimeter with the trigger
   scintillators is shown.}
   \label{kas}
\end{figure}

The investigations are performed with the KASCADE experiment,
mainly using its central detector \cite{klages, engler}.
Figure~\ref{kas} shows a sketch of the experimental arrangement.
An enlarged view of the calorimeter with its
iron absorber slabs and eight layers of liquid ionization
chambers is shown in the inset.
The third absorber gap houses 456 scintillators (0.45 m$^2$ each)
of the trigger layer.
The central detector is surrounded by a scintillator array of field
stations registrating the electromagnetic and
muonic component of EAS.

For triggering the readout of all KASCADE components,
a coincidence of at least nine scintillators of the trigger layer
is demanded. Specifically, the signals have to be above 1/3
of that of a minimum ionizing particle.
This trigger multiplicity has been proven to be an appropriate
compromise between sensitivity at low primary energies
on the one hand and reasonable permanent data amount on the 
other hand.
Given such a trigger,
in addition at least one reconstructed hadron with a minimal
energy deposit corresponding to
\mbox{$\simeq$90 GeV} is requested in the analysis of
the calorimeter. For hadrons of these energies, the reconstruction
efficiency exceeds \mbox{95 \%}.
 Typically, the trigger is generated by muons with energies
\mbox{$E_{\mu} \ge$ 0.5 GeV},
but in case 
of hadrons, cascading in the calorimeter, these also
contribute to the required trigger multiplicity.
At low primary energies
they even dominate the trigger
rate. 

The experimental observables investigated in the present studies
are two rates, i.e.~the frequency of events which fulfill
the described trigger condition, defining the {\it trigger rate},
and the frequency of events with additionally at least
one reconstructed hadron, defining the {\it hadron rate}.

The measured rates show long-term stability on the
percentage level after being corrected for dead-time and air pressure
effects (each correction $<$10 \%). 
The scintillation counters are monitored permanently utilizing the 
single muon peak, and the high voltage is corrected if necessary.
The amplification of the ionization chamber signal is checked every
six months and proves to be stable within \mbox{1.5 \%} over a
period of 4 years.
No electronic cross-talk is observed in the ionization chambers.

Uncertainties concerning the trigger layer affect also the hadron rate
being a dependent quantity.
It should be emphasized that the knowledge of the absolute energy scale
of hadrons, usually extracted from detector simulations and applied
in the same way for measurement and simulations, is not
necessary since the comparison of both data
is performed on the basis of the detector signal, i.e.~choosing
the same energy deposit in a reconstructed hadron track.
The total systematic uncertainty of the experimental values,
including threshold uncertainties, dead
or noisy channels etc.~and taking into account the steep trigger
multiplicity and hadron energy spectra, is estimated to \mbox{5 \%} for the 
trigger rate and \mbox{10 \%} for the hadron rate as listed in
\mbox{table \ref{tab.errmeas}}.
A detailed discussion of the error budget is given in 
reference \cite{risse}.

Already with data of a few days
statistical errors of the trigger and hadron rate
become negligible ($<$1 \%). 

\begin{table}\centering
 \caption{List of systematic uncertainties of the measured rates. 
          Statistical errors are negligible.}
 \label{tab.errmeas}

 \begin{indented}

  \item[] \begin{tabular}{@{}lcc}
  \br
  Source of error & trigger rate & hadron rate \\
  \mr 
  Dead-time correction & 2 \% &  2 \%  \\ 
  Air pressure correction & 1 \% & 1 \%  \\
  Trigger layer response  &  &  \\
  ~~(quadratic sum of other errors) & 4 \%  & 4 \%  \\
  Calorimeter response &  &  \\
  ~~(quadratic sum of other errors) &  &  $<$9 \% \\
\\
  Total uncertainty (quadratic sum)  &   5 \%   &  10 \% \\
  \br
  \end{tabular}

 \end{indented}

\end{table}

\section{Simulation studies of the rates}

The two observed rates are highly inclusive quantities,
which means an integration over the energy spectrum of primary
particles, over the mass spectrum, the angle-of-incidence,
and the core distance distribution of the contributing EAS.
Assuming a primary spectrum and a mass composition, the rates
can be calculated for each hadronic interaction model
by EAS simulations and by subsequently folding with the
detector response and reconstruction efficiency.
In this way, the predicted rates can be directly compared with
the measurement.

The EAS simulations are performed by use of the CORSIKA code,
and the detector response is calculated with the GEANT
package \cite{GEANT}. Five primary particle groups $-$ p, He, O,
Mg, and Fe nuclei $-$ are simulated in accordance with spectra 
obtained by
direct measurements \cite{wiebel}. For extrapolations to higher
energies \mbox{($>$500 TeV)}, the individual spectra are assumed to
drop with constant spectral indices up to a {\it knee}
energy of \mbox{$\lg (E_0/$GeV) = 6.5} and to steepen by 0.3 beyond the
{\it knee}. The complete acceptance in terms of primary energy,
zenith angle, and distance of the shower impact point to the 
central detector has been considered.
More specifically, the simulations have been performed with
primary energies of \mbox{$\lg (E_0/$GeV) = 2.5$-$7.5}, 
zenith angles of up to $45^{\circ}$ and distances of up to
\mbox{100 m}. For the highest
primary energies, angle-of-incidences up to $60^{\circ}$ and,
due to the flat muon lateral distribution, distances 
up to \mbox{500 m} from the
impact point to the central detector have been taken into account.
The core positions were scattered uniformly on the chosen area.
The maximal ranges were checked to be sufficiently large:
About \mbox{95 \%} of the contributing events are contained and a
correction for the missing fraction has been
applied.
For each model the simulation statistics corresponds
to a real-time flux of about 20 min.

CORSIKA offers
different advanced hadronic interaction models.
The calculations have been performed with QGSJET \mbox{(CORSIKA 5.62)},
\mbox{SIBYLL 1.6}, \mbox{VENUS 4.12},
\mbox{DPMJET 2.4}, \mbox{HDPM (CORSIKA 5.62)}, 
and \mbox{{\sc \normalsize neXus} 2}.
QGSJET, VENUS, and DPMJET are based on the Gribov-Regge theory
\cite{regge}.
The {\sc \normalsize neXus} model is based on a unified approach
and is currently under development in a combined effort including the
originators of QGSJET and VENUS.
Hadrons below \mbox{80 GeV} are treated by the GHEISHA \mbox{code 
\cite{fesefeldt}.}

\begin{figure}\centering
   \epsfig{file=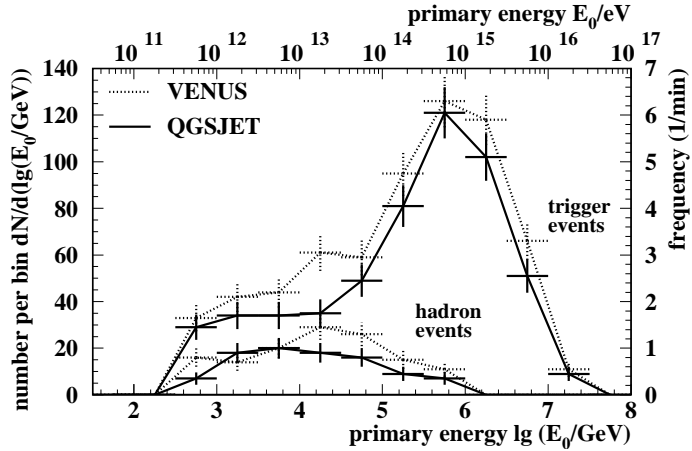,width=0.73\textwidth}
   \caption{Number of simulated events contributing to the trigger and
   hadron rate for a real-time flux of \mbox{20 min}
   vs.~primary energy. The corresponding frequency is attached to the
   right ordinate. Simulations with the models 
   VENUS and QGSJET are shown, the errors are statistical.}
   \label{eprim_spek}
\end{figure}

Figure~\ref{eprim_spek} displays the 
contributions to the trigger and hadron rate
as a function of primary energy for two different models.
Both models indicate
that the hadron rate pertains to energies below 
\mbox{1 PeV}. The trigger rate, however,
contains significant contributions also from energies
above \mbox{1 PeV}. A closer inspection shows that the hadron events are
mainly caused by proton and $-$ to a smaller extent $-$ by helium
primaries, while for the trigger rate at energies above \mbox{1 PeV}
also heavy primaries become important. 

The uncertainties of the simulated rates are listed in
table \ref{tab.errsim} and will be discussed in the following.
The main contribution to the systematic error 
stems from the uncertainty in absolute fluxes of the primary particles.
This results in a positive correlation of the systematic uncertainties 
in both rates.
For an error estimation it is favourable to distinguish the
primary energy contributions below and above
\mbox{lg($E_{0}$/GeV) = 5.5}.
Approximately, half of the trigger events and most of the hadron events
are generated in the low-energy regime. As a guideline, the 
aforementioned uncertainty of \mbox{15 \%} at \mbox{10 TeV}
\cite{watson,wiebel} is adopted.
The trigger rate requires extrapolations to high primary energies.
At \mbox{1 PeV}, e.g., the uncertainty in the absolute flux amounts to
about \mbox{40 \% \cite{watson,wiebel}}. Without changing the
absolute flux, even extreme assumptions on the 
composition or adopting a rigidity dependent {\it knee}
influence the trigger rate only little ($<$10 \%) because the 
trigger efficiencies
for different primary masses are similar to one another in the PeV region.
It should be emphasized that changes of the rates caused by an
assumption of a rigidity dependent {\it knee} are included in the
uncertainties of total flux and composition quoted in table \ref{tab.errsim}.

Since the low-energy hadronic interaction model plays an important role
for generating low-energy muons \cite{engel}, the corresponding
systematic uncertainty of \mbox{$\simeq$5 \%} has been estimated
by replacing the GHEISHA code with the program UrQMD \cite{urqmd}.
The uncertainty when simulating hadronic interactions for
the detector response is quoted separately 
to distinguish between air shower and detector simulation effects.
Other small uncertainties, like
correction for the missing fraction of contributing events, are summed up.
For a detailed list, see \cite{risse}.
The total systematic error amounts to about \mbox{25 \%} for the 
trigger rate and \mbox{20 \%} for the hadron rate.
To first approximation, however, the predictions revealed by any hadronic
interaction model would be affected in the
same manner by the mentioned systematics. 

\begin{table}\centering
 \caption{List of uncertainties of the simulated rates.}
 \label{tab.errsim}

 \begin{indented}

  \item[] \begin{tabular}{@{}lcc}
  \br
  Source of error & trigger rate & hadron rate \\
  \mr
  (a) Systematics   & & \\
  ~~~~~Primary flux, lg($E_{0}$/GeV) $<$ 5.5   &    8 \% & 15 \% \\
  ~~~~~Primary flux, lg($E_{0}$/GeV) $\ge$ 5.5: & & \\
   ~~~~~~~$-$ total flux   &  20 \% & 5 \% \\
   ~~~~~~~$-$ composition  & 10 \% & 5 \% \\
  ~~~~~Low-energy hadronic interaction in EAS  &   5 \%  &  5 \% \\
  ~~~~~Hadronic interaction in detector simulation: & & \\
   ~~~~~~~$-$ trigger multiplicity   &  5 \%  &  7 \%  \\
   ~~~~~~~$-$ total energy deposit in hadron track  &  &  4 \% \\
  ~~~~~Quadratic sum of other errors & 5 \% &  5 \% \\
\\
  Total systematic uncertainty (quadratic sum)  &  25 \%   &  20 \% \\ \\

  (b) Statistics             &  4$-$5 \%    &   8$-$10 \% \\
  \br
  \end{tabular}

 \end{indented}

\end{table}

\section{Integral rates}

\begin{figure}[b]\centering
   \epsfig{file=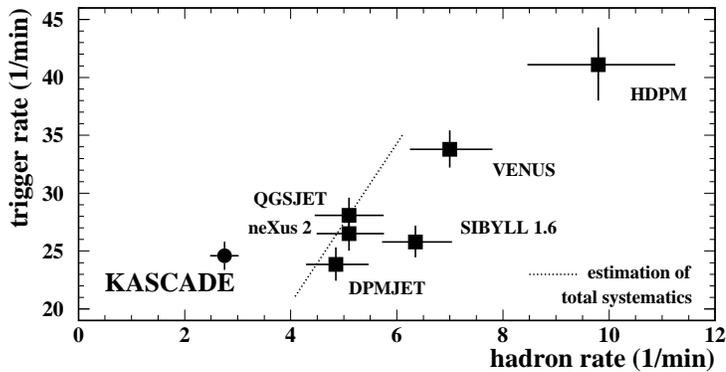,width=0.74\textwidth}
   \caption{Trigger rate vs.~hadron rate. Plotted are the measurement
   and simulated values for various models. The KASCADE value is displayed
   with the total, mainly systematic uncertainty. For simulations only
   statistical errors are shown. For the particular case of QGSJET,
   the estimated total systematic
   uncertainty (see table \ref{tab.errsim}) is indicated by the dotted line.
   The relative systematic uncertainty is the same for
   all simulations.}
   \label{raten.mod}
\end{figure}

The integral rates of all models are compiled in figure~\ref{raten.mod}
together with the experimental KASCADE value. Plotted is the trigger
rate versus the hadron rate.
The total systematic uncertainty of the predictions as listed in
table \ref{tab.errsim} is indicated for QGSJET
by the dotted line. Comparing the calculated results,
differences of about a factor 1.7 in the
predicted trigger and of 2 in the hadron rate occur. 
These discrepancies between the calculations do persist even if, for
example, different primary flux parameters
were assumed. 
All predictions would be shifted in a similar way. Thus, the scatter of
the model predictions originates from different realizations of 
the hadronic interactions in EAS.

Compared to the measurement, QGSJET, {\sc \normalsize neXus 2},
DPMJET, and \mbox{SIBYLL 1.6} predict
reasonable values for the trigger rate. On the other hand, however, all
models overestimate the hadron rate considerably.
Varying the required trigger multiplicity or hadron energy threshold does not
affect this discrepancy significantly.
The ratio of hadron rate to trigger rate of the predictions is twice the
experimental value. In this ratio, the primary flux uncertainty, e.g.,
cancels to a large extent.

\mbox{SIBYLL 1.6} is known to produce too small muon
multiplicities at observation level for primaries of PeV energy \cite{joerg},
which will be improved in the currently
developed version 2 \cite{engel}.
The reason for the \mbox{SIBYLL 1.6} prediction to agree with the measured
trigger rate is twofold: Firstly, the muon deficiency is less pronounced at lower
primary energies which also induce trigger events. Secondly, hadrons also
contribute to the trigger rate, and the overestimation of hadron events compensates
for the underestimated muon contribution.
In accordance with results in reference \cite{joerg} the larger ratio of hadron
rate to trigger rate obtained with \mbox{SIBYLL 1.6} compared to those of QGSJET
and VENUS indicates a disbalance between the hadronic and muonic component.
The correlation of hadron and trigger rate with respect to model modifications
will become apparent in the following and has been checked with a preliminary
version of \mbox{SIBYLL 2} \cite{risse}.

The differences of the simulated rates can be understood in terms of
the predicted muon lateral distribution and the number of high-energy
hadrons both of which are closely connected to the
inelastic cross-section and the elasticity distribution
of the high-energy hadronic
interactions. The large
cross-sections and low elasticity values in DPMJET, e.g.,
entail early developing
showers. Consequently, the hadron number at observation level is reduced
and the muon lateral distribution is flat with low muon densities in
the central part of the shower
which contributes most to the trigger rate. Thus, small values for
the rates emerge, and vice versa for HDPM.

To obtain an idea how the model
parameters have to be changed quantitatively to accomodate simulations
to the measurement, several changes in QGSJET
have been tried.
QGSJET has been chosen because independent analyses have shown that
this model provides the best overall description of EAS data
\cite{joerg, wolfendale}.
It turned out that increasing the inelasticity for
non-diffractive hadron-nucleus interactions 
has minor influence on
the calculated hadron rates. Most of the events 
which contribute to the hadron rate 
correspond to rare fluctuations of the free-path
of the leading hadron with very few interactions, mostly
of diffractive-dissociation type, before reaching the ground.
Consequently, the calculated hadron rates should decisively
depend on the inelastic cross-section
and on the probability for the interaction to be
diffractive. Therefore, the following (energy independent) modifications
in QGSJET have been studied in more detail:
Either the inelastic cross-section was increased 
or the fraction of diffraction dissociation was lowered keeping the
inelastic cross-section constant.
Thus, in the latter case the reduction of the diffractive 
inelastic cross-section 
is compensated by increasing the non-diffractive inelastic cross-section.
Theoretically, in the Glauber-Gribov approach \cite{glauber} one has
some parameter freedom to adjust the diffraction dissociation
cross-section independently of the inelastic cross-section without 
violating unitarity.

\begin{figure}\centering
   \epsfig{file=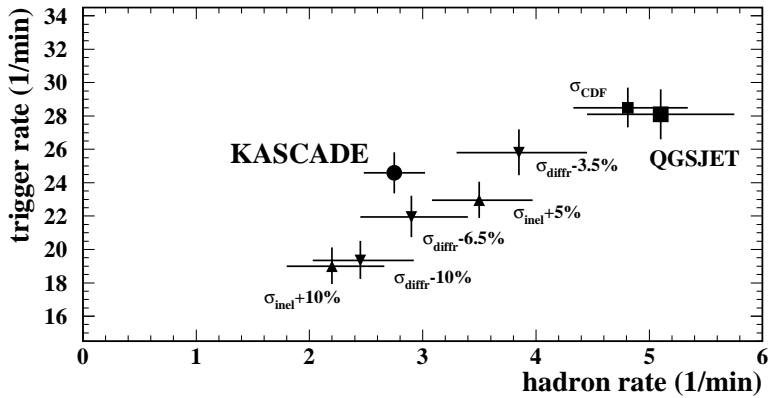,width=0.79\textwidth}
   \caption{
   Trigger vs.~hadron rate for the QGSJET model compared to
   the KASCADE measurement. Next to the
   original model prediction, the
   rates for the CDF cross-section of \mbox{80 mb} can be seen. Based
   on this version, the inelastic cross-section is increased or the
   diffraction dissociation reduced as assigned (see text).}
   \label{raten.qgs}
\end{figure}

The results are presented in figure~\ref{raten.qgs}.
To start with, the rates are shown adopting the slightly higher
CDF value for the total
proton-antiproton cross-section 
by properly changing the energy-dependent term of the Pomeron-nucleon
coupling.
The increase from
the original \mbox{76 mb} in QGSJET to \mbox{80 mb}
at \mbox{1.7 PeV} transforms to an
increase of only \mbox{1$-$2 \%} of the inelastic 
proton-air cross-section in the
energy region which is important for the hadron rate.
More specifically, the cross-section in QGSJET amounts to
\mbox{$\sigma_{inel}$(p-air) = 317/318 mb} at \mbox{10 TeV}
and \mbox{$\sigma_{inel}$(p-air) = 385/391 mb} at \mbox{1 PeV}
before/after the modification.
In fact, as can be seen from figure~\ref{raten.qgs}, no
significant effect on the rates can be stated. Additionally, 
two values with an inelastic hadron-air cross-section increased 
arbitrarily by
\mbox{5 \%} and \mbox{10 \%} and three values with diffraction
dissociation lowered by \mbox{3.5 \%}, \mbox{6.5 \%}, and \mbox{10 \%}
of the inelastic cross-section are presented.
In the original model, diffraction dissociation accounts for 
\mbox{12 \%} of
the inelastic cross-section. This number comprises all proton-nitrogen
collisions with the highest-energetic baryon carrying
more than \mbox{85 \%} of the initial momentum.
Thus, when reducing the diffraction dissociation by
\mbox{10 \%} of the inelastic cross-section, highly elastic events are
suppressed to a large extent.

One observes a large sensitivity of the rates to these changes.
Increasing the inelastic 
cross-section by \mbox{10 \%} reduces the hadron rate by
about a factor of 2.
The above-mentioned correlation of hadron and trigger
rate due to the model changes can be seen.
Additionally,
both types of modifications influence the rates in the same
way showing that the hadron rate essentially depends on the
non-diffractive part of the inelastic cross-section.
The simulations 
agree best with the measurements if the fraction of diffraction
dissociation is reduced by about a factor of 2 or, correspondingly,
by \mbox{6.5 \%} of the inelastic cross-section.

The trend of the points in figure~\ref{raten.qgs} indicates, however, that
the considered changes are not completely sufficient to 
reconcile the simulations with the measurements.
Allowing for energy-dependent modifications could be an option, but, 
as will be discussed below, also amendments with respect to
the predicted hadron multiplicities
seem necessary which might influence the calculated integral rates.

To meet the measurement given in figure~\ref{raten.qgs} by increasing
only $\sigma_{inel}$(p-air) and leaving the relative diffraction
contribution untouched, would require values
hardly compatible with results obtained at accelerators.
Both proton-antiproton cross-sections and the
elastic scattering slope are fixed by the collider data at low energies
and are already adjusted to maximum values
at \mbox{$\sqrt{s}= 1.8$ TeV}.
Using realistic equivalent radii of the nucleon distribution
in air nuclei instead of the
adopted parametrization in QGSJET might yield a \mbox{5 \%} increase of
$\sigma_{inel}$(p-air) \cite{kalmykov}.
However, such an enlargement,
still insufficient to achieve the
desirable agreement between calculations and KASCADE measurement, 
is already extreme
because the nucleon electromagnetic form factor in the
measured charge distributions has to be accounted for
\cite{batty}. 

More uncertain than \mbox{$\sigma_{inel}$(p-air)}
is the fraction of diffractive events, and we 
conclude that the
diffraction dissociation is 
overestimated in the calculated nucleon-nucleus interactions.
In case of QGSJET, diminishing the diffraction dissociation by about
\mbox{5$-$7 \%} of $\sigma_{inel}$(p-air)
seems appropriate, which corresponds to reducing
the original contribution by a factor of 2.

\section{Energy dependence}

The hadron rate is an observable highly sensitive
to the non-diffractive inelastic cross-section.
One should keep in mind the inclusive character of this quantity
with contributions of events with very different shower parameters.
For most events, generated by primary energies below the reconstruction
threshold of the KASCADE array ($\simeq$0.5 PeV),
the usual reconstruction of 
shower parameters \cite{lateral} is not possible.
However, to determine roughly the contribution of
different primary energy intervals
to the measured hadron rate, 
the muon detectors of the array are included in the analysis. 
Depending on the number of fired muon detectors,
the integral rates are grouped into different bins of
0$-$1, 2$-$3, 4$-$10, and $>$10 detectors.
A larger number of hit detectors indicates, on average, a higher primary
energy. Applying the same binning for measurement and simulation,
hadrons from the simulated events are found to
originate from primary energies typically
below \mbox{30 TeV} if not more than one muon is registered.
Demanding signals in two or three array muon detectors
selects showers around \mbox{100 TeV}, et cetera.

\begin{figure}[t]\centering
   \epsfig{file=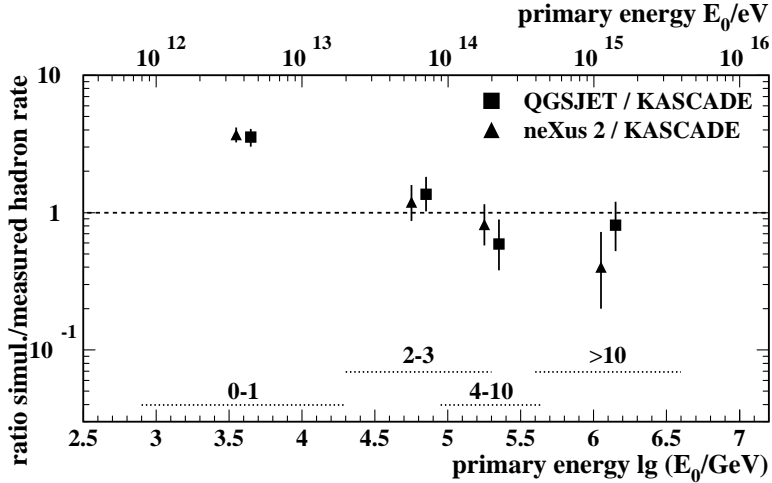,width=0.82\textwidth}
   \caption{Ratio of simulated to measured hadron rate
   vs.~energy of the
   primary particle. The simulations are performed with QGSJET
   (CORSIKA 5.62) and \mbox{{\sc \small neXus 2}}. 
   The rates are divided in four intervals depending on the 
   number of fired muon detectors as assigned (see text).
   The corresponding
   rms-width of the energy intervals is indicated at the bottom.}
   \label{array.raten}
\end{figure}

Following this scheme,
the ratios of simulated to measured hadron rates in each interval 
are given in figure~\ref{array.raten} for QGSJET and {\sc \normalsize
\mbox{neXus 2}}.
The abscissa displays the corresponding primary energies as
extracted from simulations.
Similar to the results given in figure~\ref{eprim_spek},
average and rms-width of the primary energy distributions
do not depend significantly on the interaction model
indicating that mainly the frequency of otherwise resembling events is
predicted differently.

In both models, the overestimation of the
hadron rate can be attributed to low primary energies. In this area, the hadrons
predicted by the calculations are in most cases the highest-energetic 
ones in the shower,
having suffered only a few and predominantly diffractive collisions.
Few interactions with minor energy loss
are plausible, because for normal shower developments
the small energy gap from the primary particle to the calorimeter
threshold of \mbox{90 GeV} would be too limiting.

To support this qualitative statement in a more
quantitative way, additional information available in simulations
is utilized.
In CORSIKA, for each hadron at observation level its
generation $n$ of the producing interaction 
within the cascade is recorded. For the most energetic shower hadron,
an average hadron energy of 
$E = 0.38^n \cdot E_0/A$ is expected
with $E_0$ and $A$ being the energy and mass of the primary particle
and 0.38 the mean
elasticity in QGSJET, i.e., the energy fraction of the most energetic
baryon to the initial energy in proton-air collisions.
The mean elasticity is to a good approximation energy
independent in the range of interest.

\begin{figure}[t]\centering
   \epsfig{file=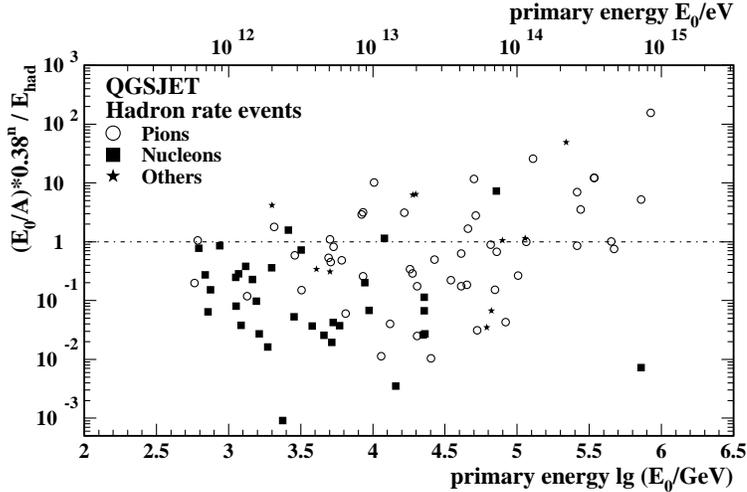,width=0.82\textwidth}
   \caption{Ratios of expected (with average elasticity)
   to actual energy of the hadron with the highest reconstructed energy
   vs.~primary energy for hadron rate events, calculated
   with QGSJET.
   Particle types are indicated as
   nucleons, pions, and others (kaons, antinucleons, ...).}
   \label{quot.lead}
\end{figure}

In figure~\ref{quot.lead} the ratio of this calculated energy to the
actual energy is plotted for the hadron with the highest reconstructed energy,
using the QGSJET model.
We observe ratios smaller than 1 at low primary
energies indicating that the
hadrons originate from highly elastic interactions. In addition,
for primaries below \mbox{10 TeV} the detected hadrons are predominantly
nucleons, the typical leading particles in proton-air collisions.
Therefore, as long as this type of events 
dominates the predicted hadron rate
one is to a large extent restricted to increase the non-diffractive part
of $\sigma_{inel}$(p-air) 
in order to reduce significantly the calculated value.
Apart from the
leading particle, details of the secondary particle
production are of minor importance in this case.

This prevalence of diffractive dissociation and of the leading particle
decreases with rising
primary energy.
Secondary hadrons in the shower
start to contribute
to the hadron rate. The ratio increases to
values well above 1, and pions, as the typical secondary particles,
become more numerous. 
Thus, at higher primary energies the hadron events can be used to test
also particle production, being in
particular sensitive to the spectra of pions and kaons
in the forward region.

\begin{figure}[t]\centering
   \epsfig{file=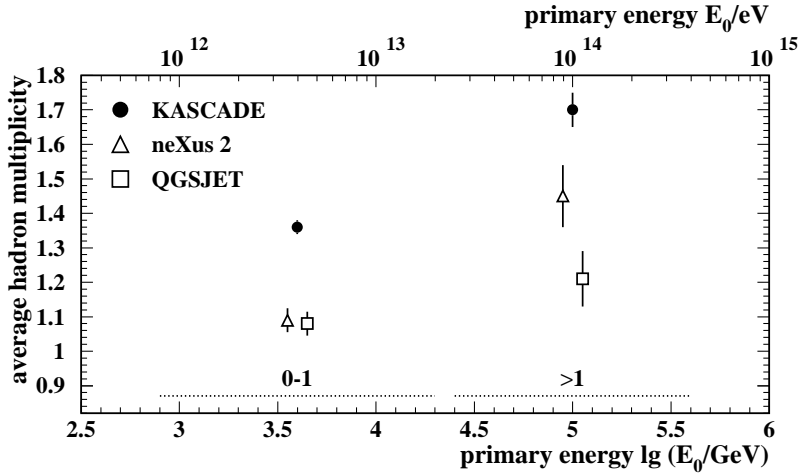,width=0.80\textwidth}
   \caption{Average multiplicity of hadrons above \mbox{90 GeV} in 
   hadron rate events vs. primary energy. Plotted are the measurement and the
   simulated values with QGSJET and {\sc \small neXus 2} for two intervals
   depending on the muon count in the array detectors as assigned (see text).
   The corresponding rms-width of the
   primary energy intervals is indicated at the bottom.}
   \label{array.hmulti}
\end{figure}

For such tests an appropriate quantity could be the mean multiplicity of
hadrons in the
calorimeter, as shown in figure~\ref{array.hmulti}. The hadron events
are grouped into intervals according to the number of muons
observed by the array, in a similar way as in
figure~\ref{array.raten}. The intervals with more than one muon are
combined to achieve a more significant result
when averaging discrete multiplicity values with limited statistics.
The gross trend, however, is apparent.
One observes a discrepancy between experiment and model
predictions at both intervals shown, especially for QGSJET. 
At low primary energies the measurement yields on average more than one
hadron, whereas in simulations in most cases
only one hadron arrives at the detector.
At higher
energies the prediction of the newly developed code {\sc \normalsize
neXus 2} agrees better with the measured hadron multiplicity.

In simulation it is checked that the multiplicity of hadrons 
reaching the detector is reconstructed correctly.
In case of multiple hadrons, both in the measurement and 
the calculations the typical separation of hadrons is much
larger than the spatial detector resolution.

The mean hadron multiplicity turns out to be quite insensitive to modifications 
of the cross-sections as given
in figure~\ref{raten.qgs} and thus allows to test additional
features of the hadronic interaction models.
To improve the agreement harder spectra of secondary pions and kaons 
appear to be required.
Future investigations, especially including the energy-multiplicity
correlation of hadrons, have to throw light onto this issue.

\section{Conclusions}

The KASCADE detector system measures observables which
allow to test high-energy hadronic interaction models
as implemented in the CORSIKA code.
The measured hadron rates are most sensitive to
primary energies between 0.5 and \mbox{500 TeV}. In this region the primary fluxes
are reasonably well known and the models are expected to be quite reliable. The
investigations reveal distinct differences in the rates between the
models which can be related to the inelastic
cross-section and the contribution of
diffractive dissociation. So far none
of the models can explain the measurements
correctly. We attribute the overestimation of the hadron rate
at small primary energies 
to an underestimate of the non-diffractive inelastic cross-section for
nucleon-air collisions. To reconcile simulations with 
measurements while keeping the inelastic cross-section unchanged,
the contribution of
diffraction dissociation should be reduced, e.g.~for QGSJET
by about \mbox{5$-$7 \%} of the inelastic cross-section. This
corresponds to about half of the original contribution.
It would therefore
be highly desirable to measure the proton inelastic cross-section
on nitrogen and oxygen at the highest energies at accelerators and
to study the kinematical region of diffraction dissociation.

Particle production in the forward region can be checked at KASCADE via
hadron multiplicities. First analyses point towards harder spectra of
secondary pions and kaons.
Further studies with the new models and in close connection to
the authors of the models are under way and 
will, hopefully, help to improve our
understanding of the hadronic interaction processes in air showers.

\section{Acknowledgments}
We are very grateful to R.~Engel and K.~Werner
for illuminating
discussions and appreciate their advice when applying their models, and
to G.B.~Yodh for carefully reading the manuscript.
The authors would like to thank the members of the engineering and
technical staff of the KASCADE collaboration who
continuously contributed to the success of the experiment.
The support of the experiment by the Ministry for Research of the German
Federal Government is gratefully acknowledged.
The Polish group acknowledges the support 
by the Polish State Committee for Scientific Research (grant No 5 
P03B 133 20). The work has partly been supported by a grant of the
Romanian National Agency of Science, Research, and Technology, by the research grant
No 94964 of the Armenian Government, and by the ISTC project A 116.
The KASCADE collaboration work is embedded in the frame of scientific-technical
cooperation (WTZ) projects between Germany and Armenia (No 002-09), Poland
(No POL-99/005), and Romania (No RUM-014-97).


\section*{References}
\begin{thebibliography} {99}
\bibitem{jones} Jones L W 1997 {\it Nucl. Phys. B (Proc. Suppl.)} {\bf 52} 103
\bibitem{kaidalov} Kaidalov A B 1979 {\it Phys. Rep.} {\bf 50} 157
\bibitem{carrol} Carrol A S {\it et al} 1979 {\it Phys. Lett.} B {\bf 80} 319
\bibitem{roberts} Roberts T J {\it et al} 1979 {\it Nucl. Phys.} B {\bf 159} 56
\bibitem{E710}   Amos N A {\it et al} (E710 Collaboration) 1992 {\it Phys. Rev. Lett.}
                 {\bf 68} 2433
\bibitem{CDF}    Abe F {\it et al} (CDF Collaboration) 1994 {\it Phys. Rev.} D 
		 {\bf 50} 5550
\bibitem{E811}   Avila C {\it et al} (E811 Collaboration) 1999 {\it Phys. Lett.} B
		 {\bf 445} 419
\bibitem{heck} Heck D, Knapp J, Capdevielle J N, Schatz G,
                  and Thouw T 1998 {\it CORSIKA: A Monte Carlo Code to
                  Simulate Extensive Air Showers} Report FZKA 6019
                 Forschungszentrum Karlsruhe;
                 http://www-ik3.fzk.de/$\sim$heck/corsika
\bibitem{watson} Watson A A 1997 {\it Proc. 25$^{th}$ Int. Conf. on Cosmic Rays (Durban)}
                 vol 8, p 257
\bibitem{klages} Klages H O  {\it et al} (KASCADE Collaboration) 1997
                 {\it Nucl. Phys. B (Proc. Suppl.)} {\bf 52} 92
\bibitem{engler} Engler J {\it et al} 1999 {\it Nucl. Instr. Meth.} A {\bf 427} 528
\bibitem{joerg}  Antoni T {\it et al} (KASCADE Collaboration) 1999 
		  {\it J. Phys. G: Nucl. Part. Phys.} {\bf 25} 2161
\bibitem{qgsjet} Kalmykov N N, Ostapchenko S S, and Pavlov A I 1997
		 {\it Nucl. Phys. B (Proc. Suppl.)} {\bf 52} 17
\bibitem{venus} Werner K 1993 {\it Phys. Rep.} {\bf 232} 87
\bibitem{sibyll} Engel J, Gaisser T K, Lipari P, and Stanev T 1992
                {\it Phys. Rev.} D {\bf 46} 5013; \\
                Fletcher R S, Gaisser T K, Lipari P, and Stanev T 1994
		{\it Phys. Rev.} D {\bf 50} 5710
\bibitem{wiebel} Wiebel B 1994 {\it Chemical composition in high energy
		 cosmic rays} Report WUB 94-08, Bergische Universt\"at
                 $-$ Gesamthochschule Wuppertal; \\
                 Wiebel-Sooth B, Biermann P L, and Meyer H 1998
                 {\it Astron. Astrophys.} {\bf 330} 389; \\
		 Apanasenko A V {\it et al} (RUNJOB Collaboration) 1999
                 {\it Proc. 26$^{th}$ Int. Conf. on Cosmic Rays (Salt Lake City)} 
		 vol 3, p 163; \\
                 Asakimori K {\it et al} (JACEE Collaboration) 1998
		 {\it Ap. J.} {\bf 502} 278
\bibitem{ranft} Ranft J 1995 {\it Phys. Rev.} D {\bf 51} 64
\bibitem{capde} Capdevielle J N 1989 {\it J. Phys. G: Nucl. Part.
                Phys.} {\bf 15} 909
\bibitem{drescher} Drescher H G, Hladik M, Ostapchenko S, Pierog T,
		   and Werner K 2001 {\it Phys. Rep.} to be
                published; preprint hep-ph/0007198
\bibitem{risse} Risse M 2000 {\it Test und Analyse hadronischer
		Wechselwirkungsmodelle mit KASCADE-Ereignisraten}
		Report FZKA 6493 Forschungszentrum Karlsruhe
	       (in German); \\
	       http://www-ik.fzk.de/KASCADE/KASCADE\_publications\_PhD.html
\bibitem{GEANT} CERN 1993 GEANT 3.21, Detector Description and Simulation
                Tool, CERN Program Library Long Writeups W5015
\bibitem{regge} Regge T 1959 {\it Nuovo Cim.} {\bf 14} 951; \\
                Gribov V N 1968 {\it Sov. Phys. JETP} {\bf 26} 414
\bibitem{fesefeldt} Fesefeldt H 1985 {\it The Simulation of Hadronic
                Showers $-$ Physics and Applications $-$} 
		Report PITHA-85/02 RWTH Aachen
\bibitem{engel} Engel R, Gaisser T K, and Stanev T 1999 {\it Proc. 29$^{th}$ Int.
                Symp. on Multiparticle Dynamics (Providence)}
		(Singapore: World Scientific) p 457
\bibitem{urqmd} Bass S A {\it et al} 1998 {\it Prog. Part. Nucl. Phys.}
		{\bf 41} 225; \\
		Bleicher M {\it et al} 1999 {\it J. Phys. G: Nucl. Part.
		Phys.} {\bf 25} 1859
\bibitem{wolfendale} Erlykin A D and Wolfendale A W 1998
               {\it Astrop. Phys.} {\bf 9} 213
\bibitem{glauber} Glauber R J 1959 {\it Lectures on theoretical physics}
		 (New York: Inter-science Publishers);  \\
		 Gribov V N 1969 {\it Sov. Phys. JETP} {\bf 29} 483
\bibitem{kalmykov} Kalmykov N N, Ostapchenko S S, and Alekseeva M K 1999
		 {\it Proc. 26$^{th}$ Int. Conf. on Cosmic Rays (Salt Lake City)} vol 1, p 419
\bibitem{batty}  Batty C J, Friedman E, Gils H J, and Rebel H 1989
		{\it Advances in Nuclear Physics} 
		vol 19 ed \mbox{J W Negele} and E Vogt
		(New York: Plenum Press), p 1
\bibitem{lateral} Antoni T {\it et al} (KASCADE Collaboration) 2001
                 {\it Astrop. Phys.} {\bf 14} 245
\end{thebibliography}
\end{document}